# Scanned probe imaging of nanoscale magnetism at cryogenic temperatures with a single-spin quantum sensor

Matthew Pelliccione[1], Alec Jenkins[1], Preeti Ovartchaiyapong[1], Christopher Reetz[1], Eve Emmanuelidu[2], Ni Ni[2] and Ania C. Bleszynski Jayich[1]*

**High spatial resolution magnetic imaging has driven important developments in fields ranging from materials science to biology. However, to uncover finer details approaching the nanoscale with greater sensitivity requires the development of a radically new sensor technology. The nitrogen-vacancy (NV) defect in diamond has emerged as a promising candidate for such a sensor based on its atomic size and quantum-limited sensing capabilities afforded by long spin coherence times. Although the NV center has been successfully implemented as a nanoscale scanning magnetic probe at room temperature, it has remained an outstanding challenge to extend this capability to cryogenic temperatures, where many solid-state systems exhibit non-trivial magnetic order. Here we present NV magnetic imaging down to 6 K with 6 nm spatial resolution and 3 µT/√Hz field sensitivity, first benchmarking the technique with a magnetic hard disk sample, then utilizing the technique to image vortices in the iron pnictide superconductor $BaFe_2(As_{0.7}P_{0.3})_2$ with $T_c$ = 30 K. The expansion of NV-based magnetic imaging to cryogenic temperatures represents an important advance in state-of-the-art magnetometry, which will enable future studies of heretofore inaccessible nanoscale magnetism in condensed matter systems.**

[1]Department of Physics, University of California, Santa Barbara, Santa Barbara, California 93106, USA

[2]Department of Physics, University of California, Los Angeles, Los Angeles, California 90095, USA

Magnetism in condensed matter systems often accompanies exotic electronic phases, and its presence is the basis for many advanced technologies. Examples include vortices in high temperature superconductors[1], ferromagnetism at the interface between oxide band insulators[2], and skyrmion phases in helimagnets[3,4]. Many experimental tools, including real space imaging techniques such as magnetic force microscopy (MFM)[5], scanning superconducting quantum interference devices (SQUIDs)[6] and Lorentz transmission electron microscopy (TEM)[7], and reciprocal space techniques including neutron scattering[8] have been successfully utilized to study magnetism in these systems. However, each of these techniques has limitations that must be considered. In MFM, a ferromagnetic tip must be placed in close proximity to a sample, which can perturb the magnetic order that is being probed. Scanning SQUIDs typically require a probe temperature of 10 K or lower, and generally offer micron-size spatial resolution, although recent studies have enhanced the resolution to submicron scales[9]. Lorentz TEM can provide images with high spatial resolution and magnetic contrast, but requires very thin samples, typically less than 100 nm thick. Neutron scattering requires the growth of large, high purity single-crystal samples, and is an ensemble-averaged measurement. There is therefore a significant opportunity to develop a real-space, non-invasive magnetic sensor capable of studying magnetic order at sub-10 nm spatial resolution and sub-$\mu$T/$\sqrt{Hz}$ DC field sensitivities.

The nitrogen vacancy (NV) defect center in diamond is an exceptionally versatile single spin system with unique quantum properties that have driven its application in diverse areas ranging from quantum information and photonics to quantum metrology[10-19]. Cryogenic scanning magnetometry stands out as potentially the most impactful application of NV centers, taking advantage of the exquisite magnetic field sensitivity and intrinsic atomic scale of the NV center for high resolution imaging[20]. The operation of an NV-based magnetic probe is dependent on a fundamentally different sensing principle than other imaging methods, namely the spin-dependent photoluminescence of a solid-state defect. Recently, the NV center has been proposed[21-23] and used[24-29] as a scanning magnetometer in a number of different contexts, but has been restricted to room temperature operation. However, NV centers maintain their high field sensitivity over a large temperature range, from cryogenic to ambient and above[15,30], and hence are ideal for imaging nanoscale magnetism through orders of magnitude in temperature. A cryogenic NV scanning magnetometer enables the study of a host of new systems that could benefit from a highly sensitive nanoscale probe, particularly solid-state systems with non-trivial magnetic order or magnetic phase transitions at low temperature.

In this work, we demonstrate the operation of an NV scanning magnetometer down to 6 K with 6 nm spatial resolution and 3 $\mu$T/$\sqrt{\text{Hz}}$ DC field sensitivity. We use the scanning NV probe to image vortices in an iron pnictide superconductor and magnetic domains in a hard disk, highlighting the compatibility of the technique with cryogenic condensed matter phenomena. We also discuss prospects for improving the spatial resolution and field sensitivity, which are limited by the separation of the NV center from the sample and coherence time of the NV center, respectively. With the coherence properties that have been demonstrated for shallow NV centers in bulk diamond[31,32], it is reasonable that with modest improvements to the scanning probes presented here, the technique can reach 3 nm spatial resolution with 500 nT/$\sqrt{\text{Hz}}$ DC and 50 nT/$\sqrt{\text{Hz}}$ AC field sensitivities.

**Single spin microscope**

A schematic representation of the system is shown in Fig. 1a. A nanofabricated single-crystal diamond cantilever with an NV center near the apex of a tip is used as the probe in a custom-built, low-temperature atomic force microscope (AFM). The cantilever consists of an array of tips containing NV centers, where a single tip is selected for imaging during a given measurement. The closed-cycle cryostat (Montana Instruments) reaches a base temperature of 4.3 K. However, in our experiment, the sample temperature is typically 6 K due to a combination of heating from the microwave excitation needed for spin manipulation, and an intentionally reduced cooling power for minimizing tip-sample vibrations. The

AFM is comprised of a six degree-of-freedom coarse positioning stage (Janssen Precision Engineering) that allows for independent alignment of the tip to the sample and of the confocal microscope optics to the NV center on the tip. A piezo scanner (Attocube ANSxyz100hs) mounted on the coarse positioning stage allows for a 9 μm sample scan range at base temperature. The AFM vibration between the tip and the sample in the vertical direction with the fridge running is 0.72 nm RMS in a 1 kHz bandwidth, measured with a tapping mode technique[33]. Long-term tip-sample drift is less than 10 nm over a period of several days, a consequence of the cryogenic environment and temperature stability of the system. The closed-cycle nature of the cryostat also has an important advantage over liquid cryogen based designs as it eliminates liquid cryogen handling, which can severely interrupt scanning probe measurements. A window-corrected objective (Olympus LCPLFN50xLCD, 0.7 NA, 3 mm working distance) that resides outside the cryostat is used for NV optical excitation and fluorescence collection, and is separated from the AFM by a pair of 200 μm thick BK7 windows, one at 300 K and the other at 40 K to act as a radiation shield. Two representative low-temperature magnetic images acquired with the NV scanning probe at $T$ = 6 K are shown in Fig. 1b,c. Individual bits on a magnetic hard drive are shown in Fig. 1b, with a bit size consistent with MFM scans taken at room temperature (Supplemental Information). Vortices in the superconductor $BaFe_2(As_{0.7}P_{0.3})_2$ ($T_c$ = 30 K) are shown in Fig. 1c. In both images, dark features trace out contours of constant magnetic field.

**Probe design and fabrication**

Figure 2a shows a scanning electron microscope image of a typical single-crystal diamond cantilever used in this work. Monolithic diamond probes were chosen over diamond nanocrystals attached to silicon AFM tips because of the superior coherence properties afforded by the bulk diamond substrate. Approximately 100 of these cantilevers were fabricated on a (100) oriented 2 mm x 2 mm diamond substrate bonded to an oxidized silicon wafer using a diamond-on-insulator approach[34]. The pillars were formed with nanoimprint lithography using a Ti hard mask and $O_2$ etching, and are approximately 200 nm in diameter, 1 µm tall and have a 1 µm pitch. The free-standing probes were then released with a deep reactive-ion etch through the silicon substrate. The density of NV centers was chosen to yield on average one NV center per pillar at a depth of 15-20 nm, with $^{14}N$ implantation performed at a dose of $5E11/cm^2$ at an energy of 15 keV, followed by 850°C annealing in vacuum and cleaning in a boiling sulfuric acid/nitric acid mixture. The rather unconventional AFM probe design with an array of tips was chosen to increase the probability of finding a shallow NV center with favorable coherence properties. The tradeoff in this design is that any given pillar may not be in direct contact with the sample surface due to geometric constraints, for example if another pillar touches the surface first. The array of pillars however could potentially provide significant imaging enhancements in a wide-field imaging scheme, where multiple pillars are measured simultaneously to image large areas quickly, and provide vector field magnetometry if NV centers of different crystallographic orientations are used.

A confocal fluorescence image of a typical diamond cantilever is shown in Fig. 2b, with 30 µW green excitation power incident on the back aperture of the objective. The relatively high fluorescence rates are a consequence of optical waveguiding by the diamond pillar structure[35], where fluorescence saturation is typically observed at incident green powers of about 1 mW. Two-photon autocorrelation measurements verified which pillars host single NV centers[36]. The diamond cantilevers are glued to a pulled glass fiber using a micromanipulator, which is then attached to a quartz tuning fork for force sensing. The tuning fork is anchored in shear mode to reduce the overall height of the tip assembly to less than 600 µm, which allows room for the cryostat window and radiation shield to fit within the working distance of the microscope objective.

**Cryogenic magnetic imaging**

We demonstrate two complementary imaging modes for the NV magnetometer shown in Fig. 3. All data in the figure were taken at 6 K on a hard disk sample. A quantitative image of the magnetic field strength at every point in the scan (a "full-field" image) can be obtained by taking the full electron spin resonance (ESR) response of the NV center at each position in the scan. The result of such a measurement is shown in Fig. 3a, where $B_{NV}$, the magnetic field along the NV axis, is plotted and obtained by measuring the frequency splitting between the two ESR peaks, $\Delta f = 2\gamma|B_{NV}|$, where $\gamma = 2.8$ MHz/G is the gyromagnetic ratio of the NV electronic spin. This splitting only sets the magnitude of the field $|B_{NV}|$; the parity of the field can be determined by applying a small external field perpendicular to the sample (± 5 G) and monitoring how the ESR splitting changes in different regions of the scan. This method has a large dynamic field range, set by the bandwidth of the ESR spectrum, but is relatively slow. In Fig. 3a, the dynamic field range during the measurement is ± 54 G, each point represents 7.5 s of averaging, and the field sensitivity is 30 µT/√Hz, which is calculated using the error in the fit of the peak positions in the ESR spectrum, along with the acquisition time per point. We note that by fitting the full ESR response, it is possible to extract more information than just $|B_{NV}|$. These parameters include the linewidth of the peaks, limited by $T_2^*$ in the low RF power limit, and the field magnitude perpendicular to the NV axis $|B_\perp|$, although this method is only sensitive to $|B_\perp|$ to second order. Measurements of $T_2^*$ are useful in decoherence imaging[37,38] where $T_2^*$ may be reduced by fast magnetic noise on the surface of the sample.

The second magnetic imaging method is outlined in Fig. 3b. This method is sensitive to a particular magnetic field contour $B_c$, which is chosen by the frequency of the RF excitation $f_{RF}$ applied during the measurement, and $B_c = \gamma(f_{RF}-f_{ZFS})$ where $f_{ZFS} = 2.878$ GHz is the NV zero field splitting at 6 K. The contour imaging method is faster than full-field imaging, and therefore has better field sensitivity at the expense of a reduced dynamic range, which is set by the width of the ESR transition and ultimately $T_2^*$. In Fig. 3b, the experimentally determined field sensitivity is 3 µT/√Hz, with 0.6 s of averaging per point. To mitigate slow fluctuations in fluorescence unrelated to the magnetic signal, the microwave excitation is cycled repeatedly on for 10 μs and off for 10 μs at each point in the scan, and the ratio is plotted. The preservation of the magnetic structure and its equivalent position in the scans shown in Fig. 3a,b, which were acquired several days apart, highlights the long term stability and noninvasiveness of the imaging system. The contour imaging method was also used to obtain the data in Fig. 1b,c.

Figure 4 demonstrates the nanoscale spatial resolution we have achieved with our NV scanning probe. Shown in Fig. 4a is a magnetic contour image ($B_c$ = 7.9 G) of a hard disk at $T$ = 6 K, and the corresponding linecut in Fig. 4b shows spatial resolution smaller than the point spacing of 6 nm. In other words, the magnetic field at two points separated by 6 nm is well resolved within the error of the measurement. When compared to the contour image in Fig. 3b, the contours in Fig. 4a are sharper due to a reduced NV-sample separation, which results in larger magnetic field gradients in the imaging plane. The spatial resolution of the imaging method, $\delta x$, is ultimately determined by the magnetic field gradient $\nabla_x B_{NV}$ and the linewidth of the NV ESR transition $\delta f = 1/T_2^*$, $\delta x = (T_2^* \cdot \gamma \nabla_x B_{NV})^{-1}$. In these measurements, $T_2^*$ = 250 ns, and $\nabla_x B_{NV} \approx$ 0.25 G/nm at the location of the contours. Importantly, Fig. 4 shows that mechanical vibrations and drift during the measurement are well below 6 nm.

Magnetometry images of vortices in the iron pnictide superconductor $BaFe_2(As_{0.7}P_{0.3})_2$ at 6 K are shown in Fig. 5a,b. The images were taken with the contour imaging method. The sample was cooled through its superconducting transition ($T_c$ = 30 K) in a 10 G magnetic field applied perpendicular to the *ab* plane of the sample. The dark circular features, which correspond to field contours of 5.9 G (Fig. 5a) and 8.8 G (Fig. 5b), are due to the stray field from the two vortices in each figure. The contours shrink in size when comparing Fig. 5a to Fig. 5b, as the NV must be brought closer to the center of the vortex to match the higher magnetic field ESR resonance condition. The measured vortex density, estimated using the larger scan area in Fig. 1c, is approximately 0.45 ± 0.1 μm$^{-2}$. This value is consistent with the calculated vortex density of $(10\ \text{G})/\Phi \approx 0.5\ \mu\text{m}^{-2}$ assuming each vortex is associated with one magnetic flux quantum $\Phi = h/(2e)$. As another confirmation of the superconducting nature of the signal, the measured magnetic response disappears when the sample is heated above $T_c$ = 30 K.

Figures 5c,d show simulations of the images in Fig. 5a,b using the computed field profile from a superconducting vortex[39] and the properties of the NV center used in the measurement. The slight asymmetry of the features, most noticeable in Fig. 5a, is well captured by the simulations, and is a result of the fact that the axis of the NV center is tilted 35.3° from the plane of the sample. Two important input parameters for the simulation are the London penetration depth $\lambda$ of the superconductor, and the NV-sample separation $h_{NV}$. If $h_{NV}$ can be determined independently, for example by imaging a magnetic reference sample[28], the NV imaging method provides a direct way of measuring $\lambda$ by fitting the results of the simulation to the contour images. Conversely, if $\lambda$ is known, $h_{NV}$ can be extracted. In our analysis, we set $\lambda$ = 200 nm based on previous measurements[40,41], which yields $h_{NV}$ = 330 nm to most closely fit the data in Fig. 5a,b.

We have presented the first NV scanning magnetometry images of superconducting vortices, and more broadly of any cryogenic condensed matter phenomena. Although vortex structure has been probed with other magnetic imaging techniques[42-44], this demonstration is important because it highlights both the high spatial resolution of NV imaging at low temperature, along with the compatibility of the NV measurement protocol with the delicate nature of the superconducting state. It is clear that neither the green excitation nor the RF excitation destroys the vortex structure, which again highlights the non-invasive nature of the technique. Given our results, it is expected that the NV scanning probe will be compatible with a wide variety of condensed matter systems at low temperature.

**Conclusion**

We have presented the first cryogenic NV scanning magnetometry images and demonstrated 6 nm spatial resolution and 3 µT/√Hz field sensitivity down to 6 K. Combining the excellent spatial resolution and sensitivity of the NV with cryogenic operation opens the door to the study of a variety of low temperature condensed matter phenomena. Future improvements to the technique will focus on extending the NV spin coherence times above 100 µs to achieve higher field sensitivities in the 10 nT/√Hz regime, along with engineering shallower NV centers to improve spatial imaging resolution down to 3 nm. In addition, an optimized design of the diamond pillars will increase photon counts via more efficient waveguiding of the NV fluorescence,[45] leading to faster acquisition times. Sample dynamics can also be probed with AC sensing protocols, which take advantage of the fast response time of the NV center to image microwave magnetic fields[18]. With these improvements, the NV scanning magnetometer is an ideal probe for a wide range of high resolution magnetic imaging studies of solid-state systems that were not possible with the established magnetic imaging toolset.

## Acknowledgements

We thank Bryan Myers, Dan Rugar, John Mamin and Bing Shen for helpful discussions. The work at UCSB is supported by an Air Force Office of Scientific Research PECASE award, DARPA QuASAR, and the MRSEC Program of the National Science Foundation under Award No. DMR 1121053. The work at UCLA is supported by the U.S. Department of Energy (DOE), Office of Science, Office of Basic Energy Sciences under Award Number DE-SC0011978. M.P. acknowledges support from the Harvey L. Karp Discovery award.

## Author Contributions

M.P. designed the experimental apparatus and carried out the experiments. M.P. and A.J. analyzed the data. A.J. and C.R. performed the simulations. P.O. fabricated the diamond probes. E.E. and N.N. provided the iron pnictide sample. M.P. and A.C.B.J. wrote the paper with feedback from all authors. A.C.B.J. supervised the project.

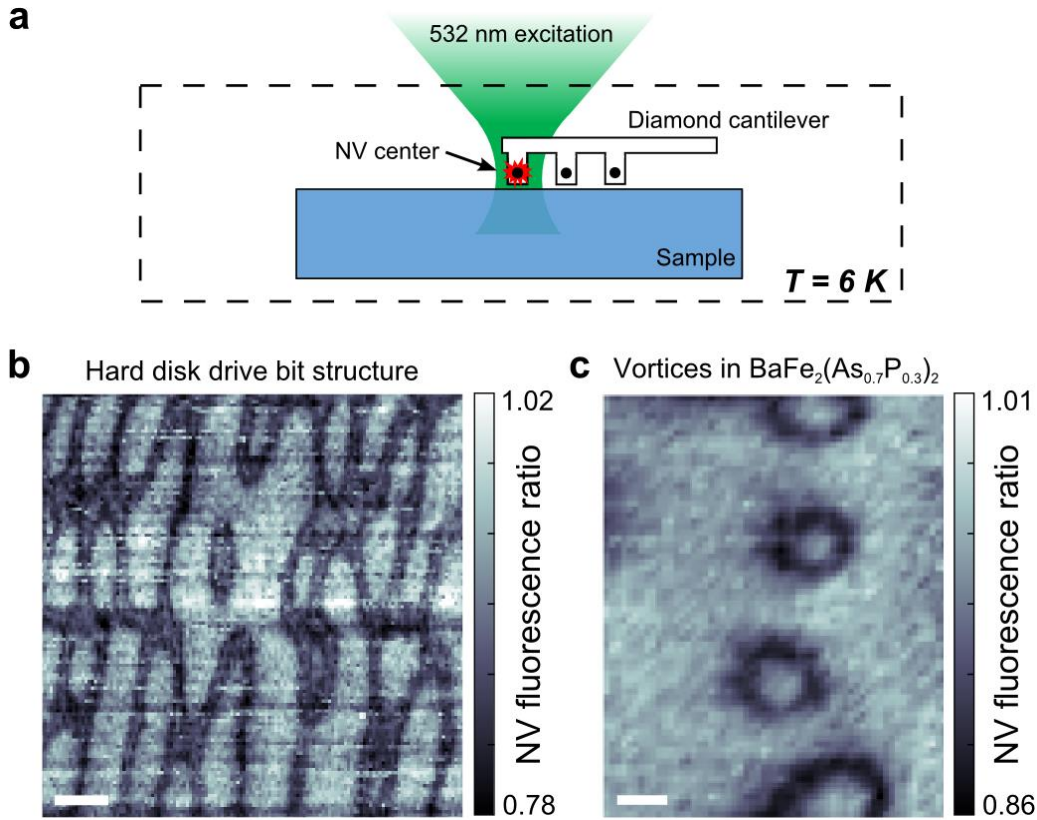

Figure 1. **Cryogenic NV scanning probe magnetometry.** (a) The scanning probe is an NV center at the apex of a tip on a diamond cantilever. The cantilever contains an array of tips where only one is selected for imaging. Both the NV center and sample are housed in a closed-cycle cryostat with a base temperature of 6 K. A 532 nm laser is focused on the NV center and the resulting fluorescence contains information about the stray magnetic field of the sample as it is scanned below the tip. Microwave excitation is applied via a gold wirebond located within 50 μm of the NV center. (b) NV magnetometry image of the bits of a hard disk at $T$ = 6 K. The dark contours in the image correspond to locations where the stray field from the hard disk has a magnitude of 5.3 G (resonant with a 2892.7 MHz RF field) along the axis of the NV center. Scale bar 100 nm. (c) NV magnetometry image of vortices in the superconductor $BaFe_2(As_{0.7}P_{0.3})_2$ at $T$ = 6 K. Analogous to (b), dark features correspond to 5.9 G magnetic field contours (2862 MHz RF field). Vortices were formed by cooling the sample through its superconducting transition ($T_c$ = 30 K) in a 10 G external field. Scale bar 400 nm.

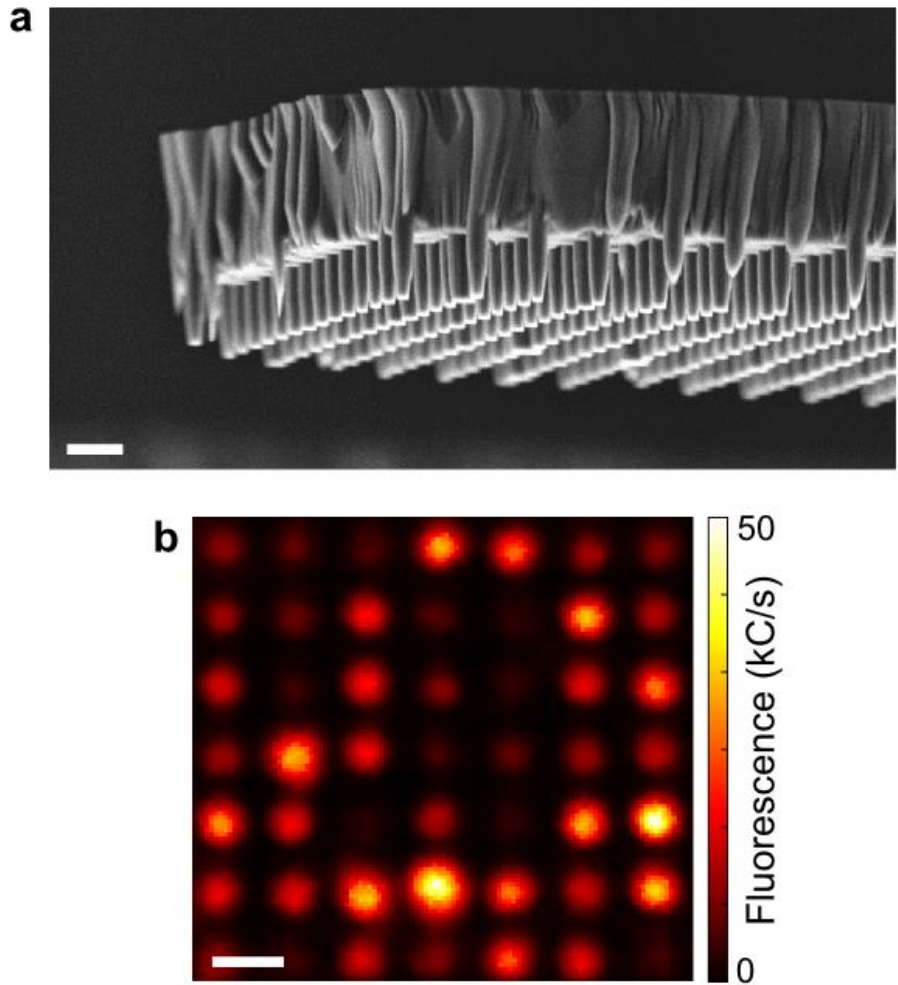

Figure 2. **Single-crystal diamond AFM probes.** (a) Scanning electron microscope image of a typical diamond cantilever fabricated for magnetometry. Each full cantilever measures 150 µm x 20 µm x 3 µm. On average, there is approximately one NV center per pillar. Scale bar 1 µm. (b) Confocal microscopy image of an array of pillars on a cantilever showing NV fluorescence from a majority of pillars. A green excitation power of 30 µW was used. Scale bar 1 µm.

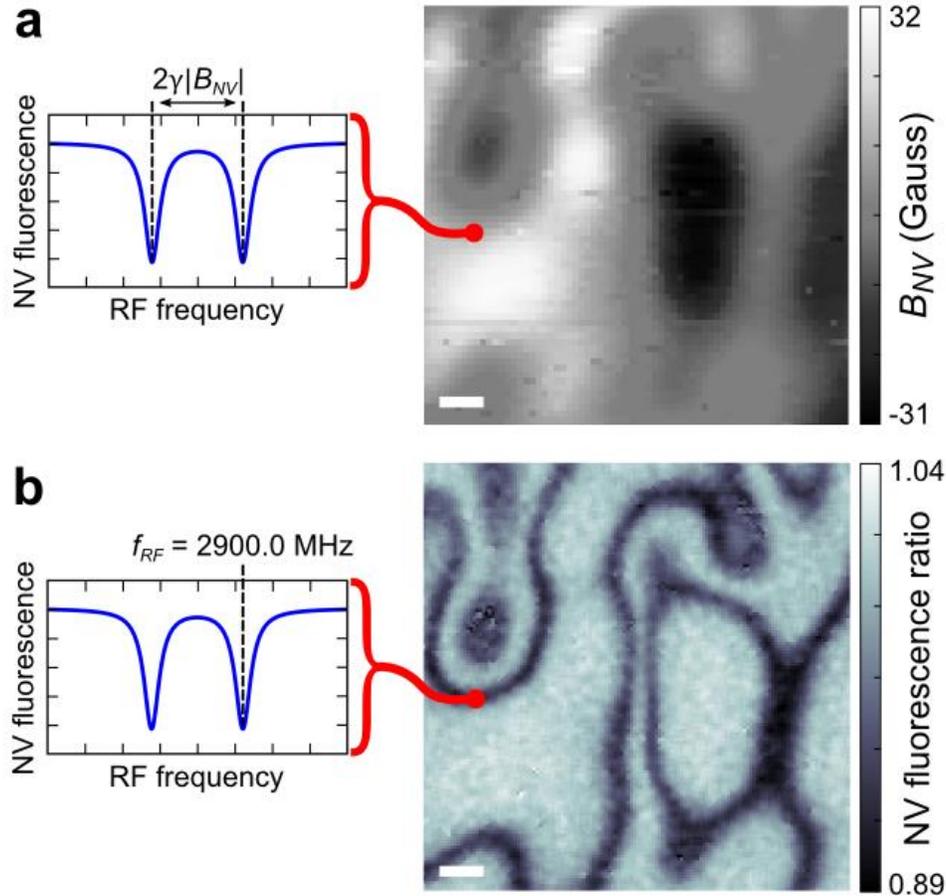

Figure 3. **Comparison of magnetic imaging protocols.** Images are taken on a magnetic hard drive at $T$ = 6 K. In (a), a full electron spin resonance (ESR) curve is acquired at each position in the scan by measuring the NV fluorescence as a function of RF excitation frequency. The image on the right shows $B_{NV}$, the magnetic field along the NV axis, where $|B_{NV}|$ is proportional to the ESR peak splitting and the sign of the field is extracted as described in the text. Scale bar 100 nm. The ESR spectrum at the position of the red dot is shown in the inset. In (b), dark features correspond to contours of constant $|B_{NV}|$ = 7.9 G, obtained when the field is on resonance with a fixed RF excitation frequency ($f_{RF}$ = 2900.0 MHz). Plotted in the image is the relative change in NV fluorescence as a result of the RF drive. Scale bar 100 nm. The images in (a) and (b) are taken over the same area of the hard disk sample.

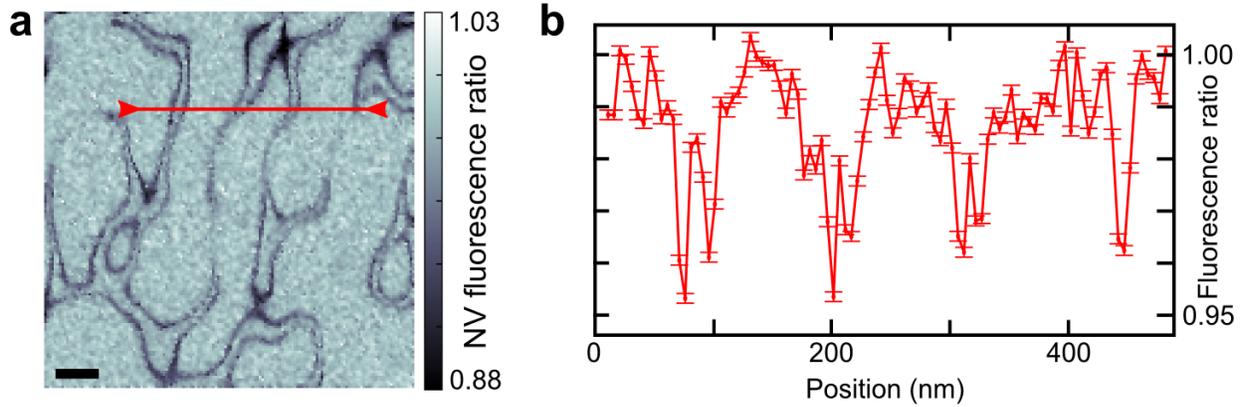

Figure 4. **High spatial resolution magnetic imaging at *T* = 6 K.** (a) Magnetic field contour image of a hard disk (7.9 G, 2900.0 MHz RF excitation). Scale bar 100 nm. (b) One-dimensional line-cut along the red line indicated on (a). Features are clearly resolved down to a spatial resolution of 6 nm, which is set by the point spacing in the scan. This data highlights the nanoscale spatial resolution of the NV sensor and the vibrational stability of the cryogenic imaging system.

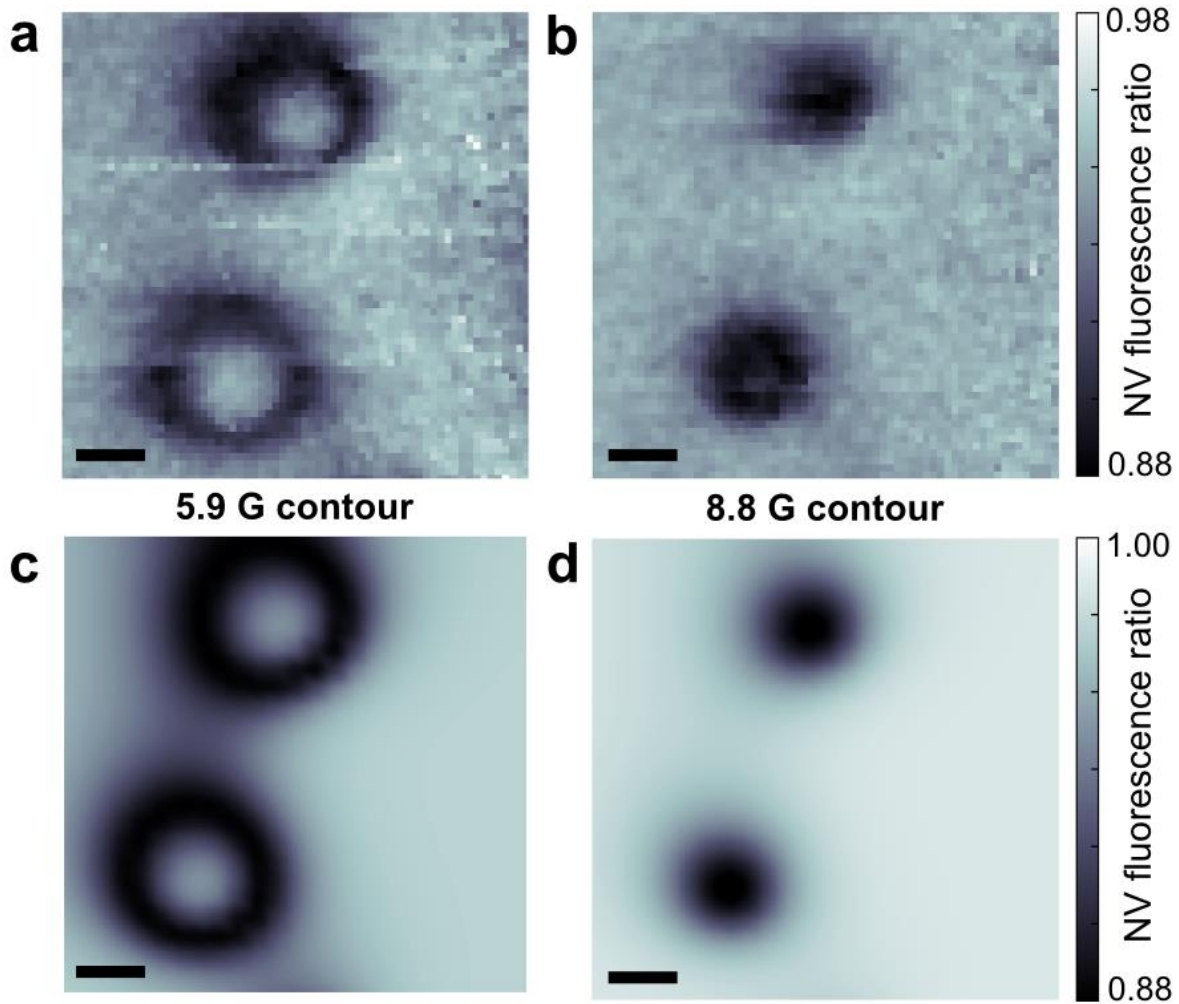

Figure 5. **NV magnetometry of vortices in the iron pnictide superconductor $BaFe_2(As_{0.7}P_{0.3})_2$ at $T$ = 6 K.** (a)-(b) Magnetic images corresponding to (a) 5.9 G and (b) 8.8 G contours of constant magnetic field. Scale bar 200 nm. The sample was cooled in a 10 G field applied perpendicular to the sample through the superconducting transition at $T_c$ = 30 K to form vortices. The dark circular features indicate the location of the vortices. (c)-(d) Simulated field contours at (c) 5.9 G and (d) 8.8 G from a vortex pair as measured by an NV center located 330 nm above the surface. Scale bar 200 nm. The London penetration depth $\lambda$ is assumed to be 200 nm in the simulation, with the NV center oriented along the [111] axis of the [100] cut diamond tip. The elongation of the vortex contours in (a) and (c) is a consequence of the stray field from adjacent vortices and the non-orthogonal orientation of the NV with respect to the sample plane.

**Supplemental Information**

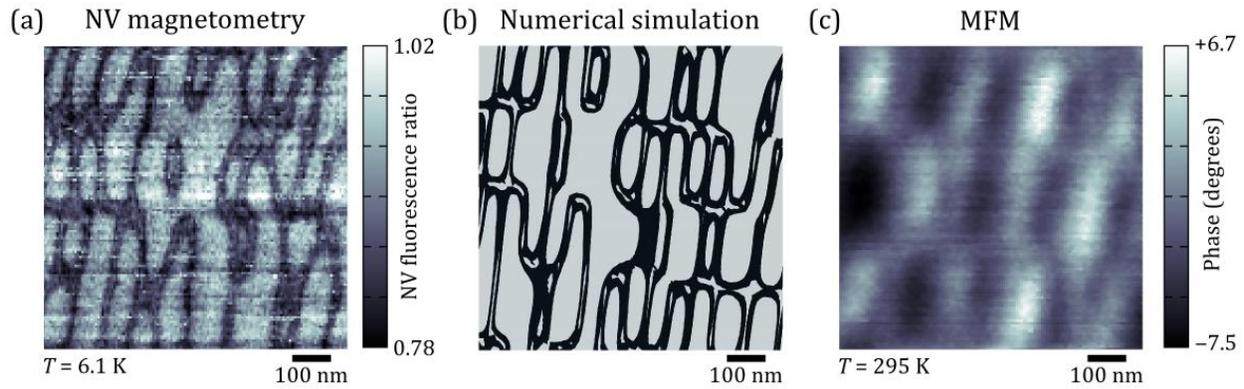

Figure S1. **Magnetic hard disk imaging at 6 K.** (a) Magnetic field contour (5.3 G) image of the hard disk at $T$ = 6.1 K. (b) Simulation of the magnetic field contours from the hard disk sample at an NV-sample separation of 73 nm. (c) Magnetic force microscopy (MFM) image of a comparable area of the hard disk at room temperature. The MFM data shows that the domains measured in (a) are consistent with the size of single bits.

**Magnetic hard disk stray field simulations**

In order to interpret the information obtained in our low temperature scans of the magnetic hard disk, we carried out simulations of the stray field produced for various arrangements of bits. Our scanning measurements were performed on a Seagate Barracuda 7200.7 hard disk. These hard disks use the older, longitudinal magnetic recording (LMR) style bits, in which the magnetization directions are parallel to the plane of the hard disk platter and are oriented along the hard disk recording tracks. The stray magnetic field from the hard disk was simulated using a two-dimensional magnetic layer with 250x100 nm bits (these dimensions were chosen based on the MFM scan shown in figure S1(c)). The magnetic structure of the hard disk was approximated by lines of effective magnetic charge, located between bits of opposite magnetization.

Figure 5(b) depicts simulated magnetic field contours from an array of bits at an NV-sample separation of 73 nm. The bit configuration only qualitatively matches the experimental contour scan in figure 5(a); there is not exact quantitative agreement. Our stray field simulations can be used to estimate the NV-sample separation in a given scan, but this technique is fairly qualitative and depends on the underlying bit pattern. For example, the bit pattern assumed in figure 5(b) contains a much higher rate of alternating bits than a truly random arrangement and as a result the stray field image contains smaller features than a random bit arrangement at the same height. After first approximating the bit pattern, we calculated the stray magnetic field along the NV axis at different NV-hard disk separations in 30 nm intervals. Comparing the resulting images to figure S1a, the NV was estimated to be 30-100 nm from the hard disk surface during that measurement.